\documentclass[prb,aps,twocolumn,showpacs,floats]{revtex4}
\usepackage{graphicx}

\begin{document}

\title{Electron structure of the Falicov-Kimball model with a magnetic field }

\author{Minh-Tien Tran}
\affiliation{Institute of Physics, Vietnamese Academy of Science
and Technology, P.O. Box 429, 10000 Hanoi, Vietnam}


\begin{abstract}
The two-dimensional Falicov-Kimball model in the presence of a
perpendicular magnetic field is investigated by the dynamical
mean-field theory. Within the model the interplay between electron
correlations and the fine electron structure due to the magnetic
field is essentially emerged. Without electron correlations the
magnetic field induces the electron structure to the so-called
Hofstadter butterfly. It is found that when electron correlations
drives the metal-insulator transition, they simultaneously smear
out the fine structure of the Hofstadter butterfly. In a
long-range ordered phase, the electron correlation induced gap
preserves the fine structure, but it separates the Hofstadter
butterfly into two wings.
\end{abstract}

\maketitle

\section{Introduction}
The problem of electrons moving under an external magnetic field
has attracted a lot of attention since the beginning of quantum
mechanics. Two dimensional electron gas under a perpendicular
magnetic field creates the quantum Hall
effect.\cite{Klitzing,Laughlin1,Tsui,Laughlin2} It comprehensively
deals with the interplay of electron correlations and magnetic
field. The integer quantum Hall effect is due to the quantization
of the energy levels of free electrons under a magnetic
field,\cite{Klitzing,Laughlin1} while the fractional quantum Hall
effect is essentially due to the electron interaction under a
magnetic field.\cite{Tsui,Laughlin2} The picture totally becomes
complexity when electrons move additionally on a lattice or under
a periodic potential. The simultaneous presence of magnetic field
and lattice potential gives the spectra of two dimensional
noninteracting electrons a fine structure of the famous Hofstadter
butterfly.\cite{Hofstadter} The Hofstadter butterfly displays a
recursive structure over rational gauge field and a Cantor set at
any irrational gauge field. The Hall conductance of the
noninteracting Bloch electrons is still quantized with an integer
number when the Fermi energy lies within a gap of the Hofstadter
butterfly.\cite{Thouless} When electron correlations are included,
the effect of simultaneous presence of magnetic field and electron
interaction on the lattice band energy remains interesting.
Without the external field electron correlations can induce
different phenomena, for instance, the metal-insulator transition,
long-range ordered phases. When electron correlations are absent,
the magnetic field creates the Hofstadter butterfly of the
electron structure. A simultaneous presence of electron
correlations and magnetic field would induce a complexity of the
electron structure. The exact diagonalization of finite lattices
shows that the electron correlations smear out the Hofstadter
butterfly.\cite{Czajka} However, it is not easy to distinguish the
fine electron structure due to the presence of magnetic field from
the discrete energy levels due to the finite-size effect of the
exact diagonalization. The Hartree-Fock mean field calculations
reveal the electron structure with the Hofstadter butterfly and an
additional correlation-induced gap.\cite{Doh,Gundmundsson} It
gives rise to an interest in the study of the interplay between
electron correlations and magnetic field in a two-dimensional
lattice beyond the Hartree-Fock approximation and in the
thermodynamic limit. In experimental aspect, with the rapid
development of ultracold technique some fundamental models of many
quantum particles can be realized by loading ultracold particles
into optical lattices (see, for example, the review in
Ref.~\onlinecite{Lewenstein}). In particular, by using the
technique of laser-assisted tunneling\cite{Jaksch} or of lattice
rotating\cite{Peden} artificial gauge fields can be created in
optical lattices. As a results it is possible to realize the
effect of magnetic field on the Bloch electrons by loading
ultracold particles into optical lattices with artificial gauge
field. Indeed, recently several models of optical lattices were
proposed to study the Hofstadter butterfly of ultracold
particles.\cite{Jaksch,Osterloh,Umucahlar,Tanatar}

In the present paper we theoretically study the effect of electron
correlations on the Hofstadter butterfly of the electron structure
under a magnetic field. The electron correlations are modelled by
the Coulomb interaction of the Falicov-Kimball model
(FKM).\cite{Falicov} It is a local repulsive interaction of mobile
electrons and massive localized particles. The FKM was originally
introduced to describe a metal-insulator transition in
transition-metal oxides. It can be viewed as a simplified Hubbard
model where electrons with down spin are frozen and do not hop.
The FKM was also used as a starting point to investigate different
electron correlation phenomena, for instance the mixed
valence\cite{Zlatic} or the electronic
ferroelectricity.\cite{Portengen,Batista} The FKM can also be
incorporated into different models to study various aspects of
electron correlations such as the charge ordered phase in
manganese compounds\cite{Ramak,Tran,Tran1} or electron
localization.\cite{Tran2,Tran3} Much progress has been made on
solving the FKM in both exact and approximation ways, where all
properties of the conduction electrons are well
known.\cite{Kennedy,Gruber1,Gruber2,FreericksZlatic} In the
homogeneous phase the FKM displays a metal-insulator transition.
When the Coulomb interaction is strong, it prevents the mobility
of itinerant electrons by forming the Mott-Hubbard gap. At low
temperature a charge ordering occurs. For half filling the charge
ordering gap opens at the Fermi level and drives the system into
an insulating phase. One may expect that the electron-correlation
induced gaps of the Mott-Hubbard type and of long-range charge
ordering may have different effects on the fine structure of the
Hofstadter butterfly. A realization of the FKM was also proposed
as an optical lattice of a mixture of light fermionic atoms (e.g.,
$^{6}$Li) and heavy fermionic atoms (e.g.,
$^{40}$K).\cite{Ates,Ziegler} When the optical lattice modelling
the FKM is established, it is also possible to create an
artificial magnetic field. In the present paper the
two-dimensional square lattice with a perpendicular magnetic field
is considered. The dynamical mean-field theory (DMFT) is employed
to calculate the electron structure of the considered model. The
DMFT is widely and successfully applied to study strongly
correlated electron systems.\cite{Metzner,GKKR} It gives the exact
solutions in infinite dimensions. However, for two-dimensional
systems the DMFT is just an approximation. It neglects nonlocal
correlations. However, the applications of the DMFT to FKM show
that the approximation is still accurate in two
dimensions.\cite{Freericks1,Freericks2} We find that the electron
correlation effect on the Hofstadter butterfly depends on the
nature of the correlated phase when the magnetic field is absent.
When a long-range order is absent, electron correlations only
induce the metal-insulator transition and they smear out the fine
structure of the Hofstadter butterfly. In a long-range ordered
phase, the electron correlation induced gap preserves the fine
structure, but separates the Hofstadter butterfly into two wings.

The plan of the present paper is as follows. In Sec.~II we
describe the FKM with a perpendicular magnetic field on a square
lattice. We also present the DMFT for calculating the Green
function in this section. In Sec.~III the numerical results are
presented. Finally, the conclusion and remarks are presented in
Sec.~IV.

\section{The Falicov-Kimball model with a perpendicular magnetic field and
the dynamical mean-field theory}

In this section we present the DMFT for the Falicov-Kimball model
in the presence of a magnetic field. The model is described by the
following Hamiltonian
\begin{eqnarray}
H &=& - \sum_{<i,j>} t_{ij} c^{\dagger}_{i} c^{\null}_{j} - \mu
\sum_{i} c^{\dagger}_{i} c^{\null}_{i} + E_{f} \sum_{i}
f^{\dagger}_{i} f^{\null}_{i} \nonumber \\
&& + U \sum_{i} c^{\dagger}_{i} c^{\null}_{i} f^{\dagger}_{i}
f^{\null}_{i}, \label{ham}
\end{eqnarray}
where $c^{\dagger}_i$ ($c_i$), $f^{\dagger}_i$ ($f_i$) are the
creation (annihilation) operators for itinerant and localized
electrons at site $i$, respectively. $t_{ij}$ is the hopping
integral of itinerant electrons between site $i$ and $j$. $U$ is
the local interaction of itinerant and localized electrons. $\mu$
is the chemical potential for itinerant electrons and $E_f$ is the
energy level of localized electrons. We will consider only the
half filling case, where $\mu=-E_{f}=U/2$. In the presence of a
magnetic field the hopping integral acquires the Peierls phase
factor\cite{Peierls}
\begin{equation}
t_{ij} = t \exp\bigg(i \frac{2 \pi}{\phi_{0}}
\int\limits_{\mathbf{R}_{i}}^{\mathbf{R}_{j}} \mathbf{A} \cdot
d\mathbf{l} \bigg) ,
\end{equation}
where $\phi_0=hc/e$, and $\mathbf{A}$ is the vector potential. For
a constant magnetic field perpendicular to the square lattice, the
Landau gauge can be chosen for the vector potential
$\mathbf{A}=(0, B x, 0)$, where $B$ is the magnetic field
strength. With this Landau gauge the hopping integral in the $x$
direction is just $t$, while in the $y$ direction it acquires
additional phase factor $t \exp(\pm i 2\pi \alpha x_{i})$, where
$\alpha=B a^2/\phi_0$, $a$ is the lattice constant, and $x_i$ is
the lattice site position in the $x$-axis. In the following we
will set $a=1$. Parameter $\alpha$ is just the magnetic flux per
unit cell in the units of the flux quantum $\phi_0$. It is clear
that the Hamiltonian is invariant with the translation $\alpha
\rightarrow \alpha+m$, where $m$ is a integer. Therefore it is
only necessary to consider $0\leq \alpha \leq 1$.

We will also only consider the rational magnetic field, i.e.,
$\alpha=p/q$, where $p$, $q$ are two coprime integers. The
translation operator that moves $q$ lattice spacing in the $x$
direction leaves the Hamiltonian unchanged. We divide the lattice
into $q$ penetrating sublattice in $x$ direction, i.e, each
lattice site can be indexed by a number $n$, and its coordinates
$x$, $y$, where $n={\rm mod}(R_{x},q)$, ($1\leq n \leq q$). We
take the Fourier transformation from the direct lattice to the
reciprocal lattice
$$
c_{n \mathbf{k}} = \frac{1}{\sqrt{N/q}} \sum_{x y} c_{n x y} e^{i
k_x x + i k_y y} ,
$$
where $N$ is the number of lattice sites. The wave vectors $k_x$,
$k_y$ are restricted to the reduced Brillouin zone
$$
-\frac{\pi}{q} \leq k_x \leq \frac{\pi}{q}, \hspace{0.5cm} \pi
\leq k_y \leq \pi .
$$
The hopping part of Hamiltonian in Eq.~(\ref{ham}) can be
rewritten as
\begin{eqnarray}
H_{t} = \sum_{\mathbf{k}} \hat{X}_{\mathbf{k}}^{\dagger}
\hat{E}(\mathbf{k}) \hat{X}_{\mathbf{k}},
\end{eqnarray}
where $\hat{X}_{\mathbf{k}}^{\dagger} =
(c^{\dagger}_{1\mathbf{k}},\ldots,c^{\dagger}_{q\mathbf{k}})$, and
\begin{eqnarray}
\hat{E}(\mathbf{k}) = -t\left(
\begin{array}{ccccccc}
\varepsilon_{1\mathbf{k}} & e^{-i k_x} & 0 & 0 & \ldots & 0 & e^{i k_x} \\
e^{i k_x} & \varepsilon_{2\mathbf{k}} & e^{-i k_x} & 0&  \ldots & 0 & 0 \\
0 & e^{i k_x} & \varepsilon_{3\mathbf{k}} &  e^{-i k_x} &\ldots &
0 & 0 \\
\vdots & \vdots & \vdots & \ddots &  \ldots & \vdots & \vdots \\
e^{-i k_x} & 0 & \ldots & & & e^{i k_x}& \varepsilon_{q\mathbf{k}}
\end{array}
\right),
\end{eqnarray}
with $\varepsilon_{n\mathbf{k}} = \cos(k_y + (n-1)2 \pi p/q )$.

We apply the DMFT to the calculation of the Green function of
itinerant electrons in the matrix form
\begin{equation}
\hat{G}(\mathbf{k},\omega) = \langle\langle \hat{X}_{\mathbf{k}} |
\hat{X}_{\mathbf{k}}^{\dagger} \rangle\rangle_{\omega} = \big[
\omega + \mu - \hat{E}(\mathbf{k}) - \hat{\Sigma}(\omega)
\big]^{-1} ,
\end{equation}
where $\hat{\Sigma}(\omega)$ is the self energy. Within the DMFT
the self energy is independent on momentum. Moreover, it is also
diagonal, i.e., $\Sigma_{nm}(\omega)=\delta_{nm}
\Sigma_{n}(\omega)$. This formulation is similar to the DMFT
applications for the antiferromagnetic or checkerboard charge
ordered phases.\cite{GKKR} Basically, the DMFT is exact in
infinite dimensions. However, its application for two-dimensional
systems is just approximation. The approximation neglects nonlocal
correlations which exist as the momentum dependence and
off-diagonal elements of the self energy. The DMFT calculations
for two-dimensional FKM without the magnetic field show that the
approximation still accurate for the electron
dynamics.\cite{Freericks1,Freericks2}

The self energy $\Sigma_{n}(\omega)$ is self consistently
determined from the dynamics of a single interaction site embedded
in an effective mean-field medium. For the FKM the effective
single-site problem can be solved exactly.\cite{FreericksZlatic}
We obtain the Green function of the single site
problem\cite{FreericksZlatic}
\begin{eqnarray}
G_{n}(\omega)=\frac{W_{n0}}{\mathcal{G}_{n}^{-1}(\omega)} +
\frac{W_{n1}}{\mathcal{G}_{n}^{-1}(\omega)-U} ,
\end{eqnarray}
where $\mathcal{G}_{n}(\omega)$ is the Weiss field for sublattice
$n$. The weight factors $W_{n0}$ and $W_{n1}$ can be calculated
from the Weiss field
\begin{eqnarray}
W_{n1} &=& f(\widetilde{E}_{n}) , \\
W_{n0} &=& 1 - W_{n1} ,
\end{eqnarray}
where $f(\omega)=1/(\exp(\omega/T)+1)$ is the Fermi-Dirac
distribution function, and
\begin{eqnarray}
\widetilde{E}_{n} = E_{f} - \int \frac{d\omega}{\pi} f(\omega)
{\rm Im} \log\bigg(\frac{1}{1-U \mathcal{G}_{n}(\omega)}\bigg) .
\end{eqnarray}
The Weiss field Green function $\mathcal{G}_{n}(\omega)$ also
satisfies the Dyson equation of the effective single site problem,
i.e.,
\begin{equation}
G^{-1}_{n}(\omega) = \mathcal{G}_{n}^{-1}(\omega) -
\Sigma_{n}(\omega) .
\end{equation}
The self-consistent condition requires that the Green function
obtained from the effective single-site problem must coincide with
the local Green function, i.e.,
\begin{eqnarray}
G_{n}(\omega) = \frac{1}{N/q} \sum_{\mathbf{k}}
G_{nn}(\mathbf{k},\omega) .
\end{eqnarray}
With this self-consistent condition the system of equations for
the self energy is closed. We can solve the system of equations by
iterations as usual.\cite{FreericksZlatic}

\section{Numerical results}

\begin{figure}[t]
\includegraphics[width=0.48\textwidth]{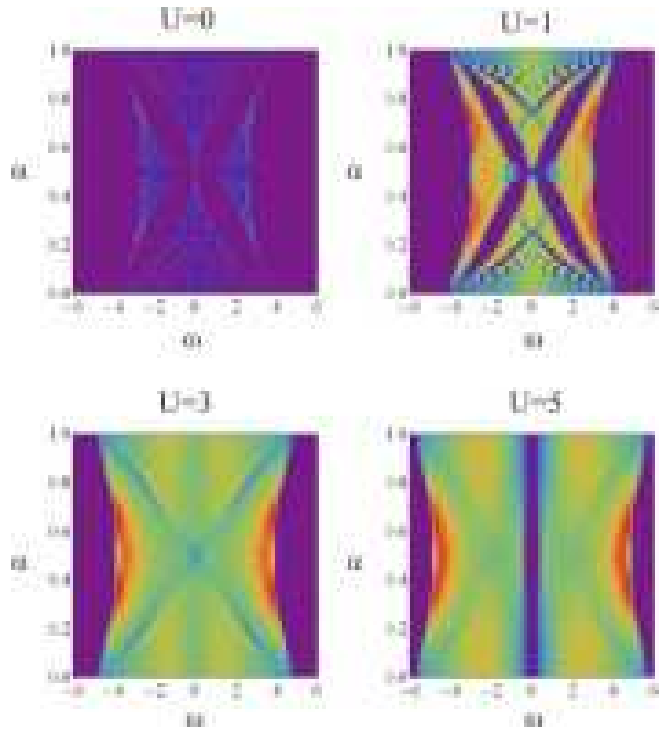}
\caption{(Color online) The density plot of the DOS of itinerant
electrons of the  high temperature phase ($T=1$) for various
values of $U$. The magnetic field parameter $\alpha=p/q$ varies
with $p=0,1,\ldots, q$, and $q=40$. The color scheme corresponds
to the rainbow color scheme, i.e. the red (violet) color
corresponds to the largest (smallest) value of the DOS.}
\label{figbuthigh}
\end{figure}

In this section we present the numerical results obtained by
solving the DMFT equations by iterations. We take $t=1$ as the
energy unit. The magnetic parameter $\alpha=p/q$ varies with
$p=0,1,\ldots, q$. We take $q=40$. When $p$ and $q$ are not
coprime, we can reduce them to coprime integers. This also reduces
the matrix dimension of the Green function, and saves the
computation time. First we consider the high temperature phase,
where the metal-insulator transition of the Mott-Hubbard type
occurs. We emphasize that in this phase temperature doest not
affect on the electron structure. The phase is just the
homogeneous solution of the DMFT equations and it is stable at
high temperature. The electron structure can be imaged by using
the density plotting of the density of states (DOS) of the
itinerant electrons. In Fig.~\ref{figbuthigh} we plot the image of
the DOS for various values of $U$. It shows that the electron
structure mimics the Hofstadter butterfly when the electron
correlations are included. For weak interactions the fine
structure of the Hofstadter butterfly still survives. However, the
electron correlations already smear out it. Some fine gaps in the
structure of the Hofstadter butterfly are closed. As the value of
$U$ increases, the smearing becomes stronger. For strong
interactions all fine gaps are closed. However, a middle rough gap
opens for $U>U_c \approx 4t$. This gap is essentially the
Mott-Hubbard gap, which opens in the insulating phase. Without the
magnetic field the metal-insulator transition occurs at $U_c$. In
the presence of the magnetic field the metal-insulator transition
still occurs, but the lower and upper bands mimic the Hofstadter
butterfly. The electron structure is symmetry in respect to lines
$\omega=0$ and $\alpha=1/2$ like the noninteraction case, i.e.,
$\rho(\omega,\alpha)=\rho(-\omega,\alpha)=\rho(\omega,1-\alpha)$,
where $\rho(\omega,\alpha)=-\sum_{n}{\rm Im} G_{n}(\omega)/q \pi $
is the DOS of itinerant electrons.

\begin{figure}[t]
\includegraphics[width=0.5\textwidth]{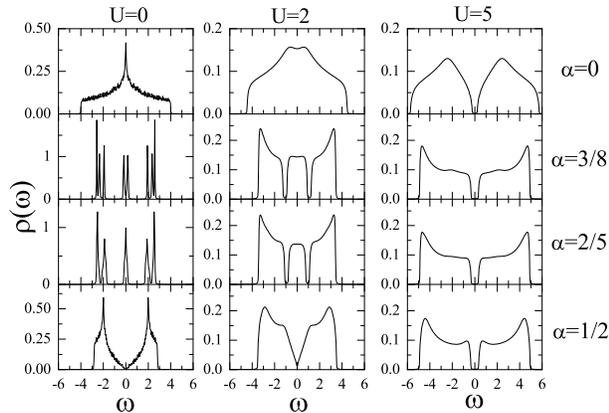}
\caption{The DOS of itinerant electrons of the high temperature
phase ($T=1$) for various values of $\alpha$ and $U$.}
\label{figdoshigh}
\end{figure}

\begin{figure}[b]
\includegraphics[width=0.48\textwidth]{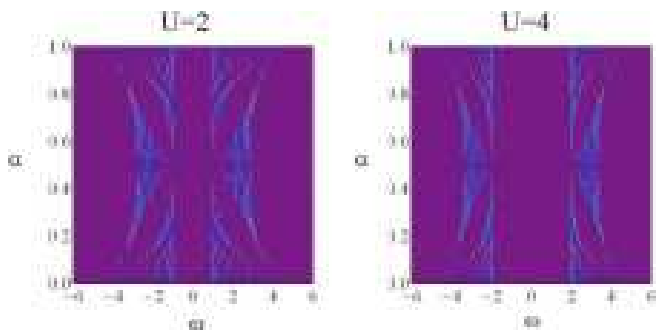}
\caption{(Color online) The density plot of the DOS of itinerant
electrons of the  charge ordered phase at low temperature
($T=0.01$) for various values of $U$. The magnetic field parameter
$\alpha=p/q$ varies with $p=0,1,\ldots, q$, and $q=40$. The color
scheme corresponds to the rainbow color scheme, i.e. the red
(violet) color corresponds to the largest (smallest) value of the
DOS.} \label{figbutlow}
\end{figure}

In Fig.~\ref{figdoshigh} we plot the DOS of itinerant electrons
for various values of $\alpha$ and $U$. When the electron
correlations are absent ($U=0$) the number of bands is just
$q$.\cite{Hofstadter} One can observe that the $q$ bands can be
grouped into subgroups of bands which are separated by moderate
gaps. Within a band subgroup the bands are also separated by fine
gaps. For example, when $\alpha=3/8$, there are $3$ subgroups of
bands, one at $\omega=0$, and two others around $\omega=\pm 2$.
When the interaction is included, the band number is reduced by
closing the gaps. As the interaction increases, first the fine
gaps within the band subgroups are closed, and then the gaps
between the band subgroups are closed. In the insulating phase all
the gaps in the Hofstadter butterfly are closed, but the
Mott-Hubbard gap opens. The metal-insulator transition occurs at
the same value of $U$ for all values of the magnetic field.
Despite of closing of the fine gaps in the insulating phase, the
intensity of the DOS of itinerant electrons still shows a mimic
Hofstadter butterfly with smearing out fine features, as shown in
Fig.~\ref{figbuthigh}.

Without the magnetic field the FKM also displays the checkerboard
charge ordering at low temperature for any interaction $U\not= 0$.
We study the possibility of the charge ordering when the magnetic
field is present. In this case we additionally divide the lattice
into two penetrating sublattices in the $y$ direction. If $q$ is
odd integer, we double the value of $p$ and $q$ that
$\alpha=2p/2q$ and the checkerboard charge ordering is
commensurate with the magnetic structure. In Fig.~\ref{figbutlow}
we plot the image of the total DOS of itinerant electrons in the
checkerboard charge ordered phase. It shows that the fine
structure of the Hofstadter butterfly still remains, however, it
is separated by a middle gap. The middle gap is the charge
ordering gap, which locks itinerant electrons into the
checkerboard charge pattern. As the interaction increases the
charge ordering gap also increases, and the band width of the
lower and upper bands are slightly reduced. The preservation of
the fine structure of the Hofstadter butterfly was also observed
within the Hartree-Fock mean field
calculations.\cite{Doh,Gundmundsson} However, the Hartree-Fock
mean-field approximation cannot find the smearing of the
Hofstadter butterfly due to electron correlations when a
long-range order is absent.

\begin{figure}[t]
\includegraphics[width=0.48\textwidth]{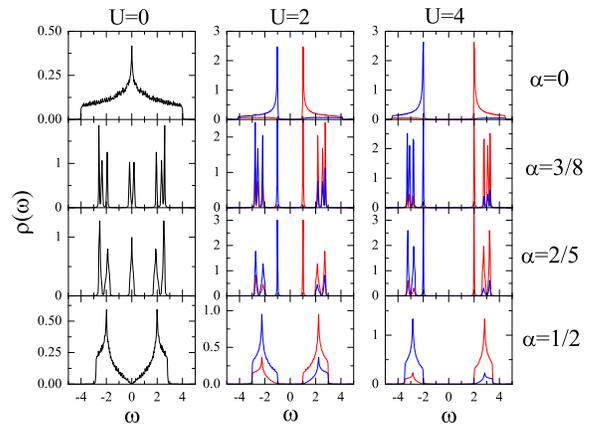}
\caption{(Color online) The DOS of itinerant electrons of the
checkerboard charge ordered phase at low temperature ($T=0.01$)
for various values of $\alpha$ and $U$. The red and green color
lines correspond to the DOS of two penetrating sublattices of the
checkerboard charge ordered phase.} \label{figdoslow}
\end{figure}

In Fig.~\ref{figdoslow} we also plot the DOS of itinerant
electrons of the two penetrating sublattices of the checkerboard
charge ordered phase for various values of $U$ and $\alpha$. It
shows that the magnetic field does not affect on the charge
ordering gap. The gap sorely depends on the interaction as in the
case of absence of the magnetic field. When $q$ is even, the
number of bands is also $q$, like the noninteraction case.
However, unlike the noninteraction case, the subgroup of bands at
the Fermi level is separated by the charge ordering gap. When $q$
is odd, the band at the Fermi level is split into two bands which
are also separated by the charge ordering gap, as shown in
Fig.~\ref{figdoslow} (the case $\alpha=2/5$). The gaps between
subgroups of bands are slightly reduced as the interaction
increases.

\section{Conclusion}

In the present paper we study the effect of electron correlations
on the Hofstadter butterfly which is the electron structure of the
two-dimensional Bloch electrons under a perpendicular magnetic
field. By employing the DMFT we calculate the Green function of
itinerant electrons in the case of rational magnetic field.
Electron correlations exhibit different effects on the Hofstadter
butterfly depending on the nature of the correlated phase when the
magnetic field is absent. In the absence of a long-range order,
electron correlations also induce the metal-insulator transition
when the magnetic field is present. However, the electron
correlations smear out the fine structure of the Hofstadter
butterfly. The number of bands is reduced as the interaction
increases. In the insulating phase, all fine gaps of the
Hofstadter butterfly are closed, but the Mott-Hubbard gap opens at
the Fermi level. In a long-range ordered phase, the electron
correlation induced gap, such as the checkerboard charge ordering
gap in the FKM, preserves the fine structure of the Hofstadter
butterfly. However, the Hofstadter butterfly is separated into two
wings by the long-range ordering gap.

In the present paper we have only considered the rational magnetic
field. In the noninteraction case an irrational magnetic field
induces the Hofstadter butterfly in the form of a Cantor set. The
magnetic structure is incommensurate with the lattice structure as
well as with the checkerboard charge ordering pattern. The effect
of electron correlations on the such Hofstadter butterfly remains
open, and we leave it for further study.

\begin{acknowledgments}

This work was supported by the Vietnamese NAFOSTED.

\end{acknowledgments}

\vspace{0.5cm}

{\bf Note}: When this paper is written up, a preprint\cite{Wrobel}
was published in which the same model was studied and similar
results were obtained by Monte-Carlo simulations on finite
lattices.

\end{document}